# Electrical spin injection into Si(001) through a SiO$_2$ tunnel barrier

C. H. Li,[1,a)] G. Kioseoglou,[1,2] O. M. J. van 't Erve,[1] P. E. Thompson,[1] and B. T. Jonker[1,b)]

[1]*Naval Research Laboratory, Washington, DC 20375*

[2] *Department of Materials Science and Technology, University of Crete, Heraklion, Crete 71003, Greece*

ABSTRACT

We demonstrate spin polarized tunneling from Fe through a SiO$_2$ tunnel barrier into a Si *n-i-p* heterostructure. Transport measurements indicate that single step tunneling is the dominant transport mechanism. The circular polarization, $P_{circ}$, of the electroluminescence (EL) shows that the tunneling spin polarization reflects Fe majority spin. $P_{circ}$ tracks the Fe magnetization, confirming that the spin-polarized electrons radiatively recombining in the Si originate from the Fe. A rate equation analysis provides a lower bound of 30% for the electron spin polarization in the Si at 5 K.

[a)] Electronic mail: cli@anvil.nrl.navy.mil
[b)] Electronic mail: jonker@nrl.navy.mil

Silicon is an ideal host for spin as an alternate state variable, because its low atomic mass and crystal inversion symmetry result in very small spin orbit interactions and long electron spin lifetimes.[1] Efficient electrical spin injection into Si using readily fabricated contacts is a prerequisite for developing spin-based functionalities in Si.[2,3] We recently demonstrated electrical injection of spin-polarized electrons from an Fe film through an $Al_2O_3$ tunnel barrier into Si (001), resulting in an electron spin polarization of ~30% in the Si at 5 K, with significant polarization extending to at least 125 K.[4,5] A tunnel barrier alleviates the conductivity mismatch between the metal and semiconductor which would otherwise prevent spin injection,[6-8] and prevents intermixing and Fe silicide formation at the metal/semiconductor interface.[9]

$SiO_2$ has been the gate oxide of choice for Si metal-oxide-semiconductor devices, because it is easy to form and provides the low interface state density necessary for device operation. However, there are few reports of its use as a spin tunnel barrier. Smith *et al.* reported a smaller than expected magnetoresistance of 4% at 300 K in $Co/SiO_2/CoFe$ tunnel junctions, which they attributed to over-oxidation at the metal interfaces.[10] A composite $NiFe/Al_2O_3/SiO_2$ tunnel barrier was used as the emitter in a magnetic tunnel transistor by Park *et al.*[11] They reported a tunnel spin polarization of 27% at 100 K, which was lower than the value of 34% obtained for a $NiFe/Al_2O_3$ emitter in the same structure.

We report here spin polarized tunneling from Fe through an $SiO_2$ tunnel barrier into a Si *n-i-p* heterostructure, producing an electron spin polarization in the Si, $P_{Si} >$ 30% at 5 K. Single step tunneling is the dominant transport mechanism at low bias. The circular polarization of the electroluminescence (EL) tracks the Fe magnetization, confirming that the spin polarized electrons originate from the Fe.

The Si samples were grown by molecular beam epitaxy (MBE) and consist of 70 nm $n$-Si / 70 nm undoped Si / 150 nm $p$-Si / $p$-Si(001) substrate to form a light emitting diode (LED) structure. The growth temperature was reduced from 800 to 500 °C for the n-layer to achieve an n-doping $\sim 4\times 10^{18}$ cm$^{-3}$ (at 300 K) so that the electrons are free rather than donor bound even at low temperature. A dilute hydrofluoric acid etch (10%) and de-ionized water rinse was used to H-passivate the air-exposed Si surface before the samples were loaded into a second MBE system. After thermal desorption of the H in vacuum (820 K), the sample was cooled to room temperature, transferred back into the loadlock, and exposed to 200 Torr of $O_2$ for 20 minutes. An *in situ* ultraviolet (UV) lamp provided an ozone partial pressure which enhanced the Si surface oxidation process, resulting in ~2nm thick layer of $SiO_2$.[12-14] The sample is transferred in vacuum into the second MBE growth chamber, where a 10 nm Fe film is deposited below room temperature, followed by a 2 nm Al cap to prevent Fe oxidation. Details of the Si growth and contact deposition are found in Ref. 4.

Transport measurements were used to evaluate the conduction process from the Fe into the top n-Si layer for each contact. A typical I-V curve at 10 K is shown as an example in Fig. 1(a), which exhibits the non-linearity around zero bias expected for a tunnel barrier. A few of the temperature dependent conductance curves are shown in Fig. 1(b). We obtain good fits (solid lines) to the conductance curves within an energy range of ±100 meV using the Brinkman, Dynes, and Rowell (BDR) model[15] which yield an effective barrier height of ~1.7 eV and a barrier thickness of ~21 Å for 10 K. The temperature dependence of the zero-bias resistance (ZBR), expressed as $R_0(T)/R_0(300 K)$, where $R_0$ is the slope of the I-V curve at zero bias, is summarized in Fig. 1(c). The ZBR exhibits a weak temperature dependence indicative of single step tunneling rather than the exponential dependence of thermionic emission. A weak temperature dependence of the ZBR has been shown to be a reliable indicator for a pinhole-free tunnel barrier.[15]

The samples were processed into surface-emitting LEDs using standard photolithography and chemical etching techniques. Individual diodes were biased to inject electrons from the Fe into the Si, and the electroluminescence emitted along the surface normal was analyzed for positive and negative helicity ($\sigma+$ and $\sigma-$) in the Faraday geometry. The circular polarization is defined as $P_{circ} = [I(\sigma+)-I(\sigma-)]/[I(\sigma+)+I(\sigma-)]$, where $I(\sigma+)$ and $I(\sigma-)$ are the intensities of the EL analyzed for $\sigma+$ and $\sigma-$, respectively.

EL spectra at T = 5 K and B = 0 and 3 T are shown in Fig. 2, analyzed for $\sigma+$ and $\sigma-$ circular polarization. The spectra are dominated by transverse acoustic (TA) and transverse optical (TO) phonon-mediated recombination in the Si, as expected. Although emission is initially detected at a bias of 1.6V, higher biases are typically used to reduce data acquisition time, as the polarization is relatively independent of the bias. The three major emission peaks are identified as TA (1.105 eV), TO-TA (1.090 eV), and TO+TA (1.05 eV) associated with the boron acceptor. Applying a magnetic field along the surface normal rotates the Fe magnetization out of plane, enabling the injected electron spin polarization to be manifested as circular polarization in the EL. At B = 3 T, where the Fe magnetization is saturated out of plane, the spectra exhibit polarization for each of the spectral features. At higher temperatures and higher biases, the TA-related features are suppressed and the TO at 1.07 eV dominates the emission.

The magnetic field dependence of $P_{circ}$ for the TA, TO-TA and TO+TA features at T = 5 K are summarized in Fig. 3. $P_{circ}$ consistently tracks the out-of-plane magnetization of the Fe (dashed line) and saturates at ~ $4\pi M$(Fe) = 2.2 T, demonstrating that the circular polarization is due to radiative recombination of spin polarized electrons which originate from the Fe. This also argues against significant Fe-O or Fe-Si compound formation at the interface due to pinholes or diffusion, which would lead to a different magnetic field dependence. The highest energy feature at each temperature exhibits the largest $P_{circ}$, consistent with spin relaxation accompanying energy loss, and is taken to be the most representative of the initial electron spin polarization in the silicon. $P_{circ}$ of the TA at T = 5 K saturates at approximately 3%. $P_{circ}$ from reference samples fabricated on the same *n-i-p* substrate with non-magnetic metal / tunnel barrier contacts exhibits a weak paramagnetic behavior of order 0.1%/T,[4,9] much smaller than the polarization observed from the Fe/SiO$_2$ LEDs. Possible contributions to the measured $P_{circ}$ from dichroism as the emitted light passes through the Fe film have also been calculated and measured with photoluminescence on an undoped reference sample, and found to be less than ±0.5% [4]. The emission intensity decreases significantly with temperature and becomes difficult to measure above 50 K, although the polarization remains the same as that at 5 K.

For comparison, Fe/Al$_2$O$_3$ tunnel barrier contacts were grown on a piece of the same Si *n-i-p* substrate, and surface emitting LEDs fabricated as described previously.[4] The EL spectra from these LEDs are shown in Fig. 4 for reference, and exhibit the same spectral features and magnetic field dependence (inset). $P_{circ}$ for the TA feature of the Fe/Al$_2$O$_3$ contact LEDs saturates at a value of ~ 2.5%. The optical polarization found in these LEDs is slightly lower than that reported previously (e.g. 3.7-3.8% for the TA,[4,9] which we attribute to the minor variations between the different Si *n-i-p* structures used.

Due to the nature of phonon-assisted radiative recombination in indirect-gap materials such as Si, the photons measured will carry only a fraction of the spin angular momentum of the initial spin-polarized electron population. Thus $P_{circ}$ provides only a lower bound for the electron spin polarization in the Si. In addition, since the electron spin decays exponentially with a characteristic time $\tau_s$ before radiative recombination, the optical polarization that we measure is further limited by the ratio of the spin lifetime $\tau_s$ and the radiative lifetime $\tau_r$, which typically is very long for indirect-gap materials. Using a rate equation analysis as described in Ref. 4 which takes these lifetimes into account, we obtain a value of 30% as a lower bound for the electron spin polarization achieved in the silicon at T = 5 - 50 K.

In summary we have demonstrated electrical spin injection into Si from an Fe/SiO$_2$ tunnel contact. Single step tunneling is found to be the dominant spin transport mechanism. This study demonstrates that an ultra thin layer of SiO$_2$, readily fabricated on Si through UV-ozone oxidation, can be used as a viable tunnel barrier for electrical spin injection from a ferromagnetic contact into Si.

Acknowledgements: This work was supported by the Office of Naval Research and core programs at NRL. GK and OvtE gratefully acknowledge support as NRL/George Washington University Research Associates.

**FIGURE CAPTIONS**

**Figure 1.** Transport measurements across the Fe/SiO$_2$/Si tunnel junction with *n*-type doping of 4x10$^{18}$/cm$^3$ in Si. (a) Typical I-V curve at 10 K. (b) BDR model fits to conductance curves at 10, 20, 30, and 50 K. (c) Temperature dependence of zero bias resistance (ZBR).

**Figure 2.** Electroluminescence spectra at 5 K from a Si *n-i-p* structure with an Fe/SiO$_2$ contact at zero field and 3 T, analyzed for σ+ and σ- circular polarization.

**Figure 3.** Magnetic field dependence of $P_{circ}$ for the TA, TO-TA, TO+TA features at 5 K. $P_{circ}$ consistently tracks the out-of-plane magnetization of the Fe (dashed line).

**Figure 4.** Electroluminescence spectra at 5 K for the same Si *n-i-p* structure with an Fe/Al$_2$O$_3$ tunnel contact at zero field and 3 T. Inset: magnetic field dependence of $P_{circ}$ for the TA, TO-TA, TO+TA features at 5 K.

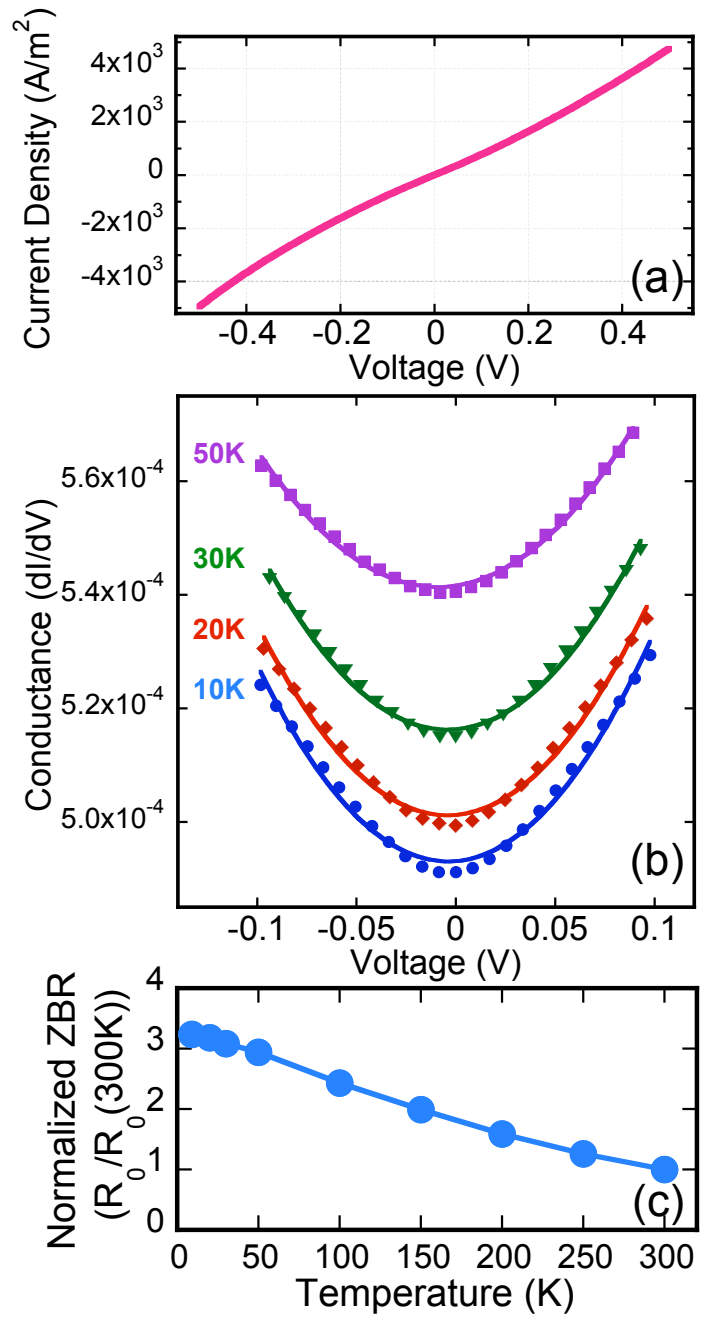

Figure 1. Li *et al.*

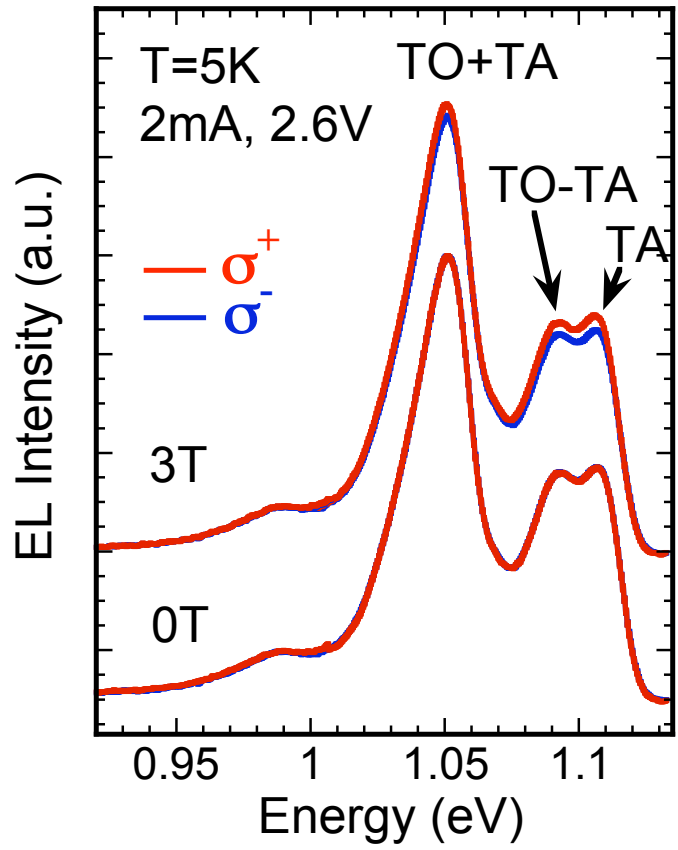

Figure 2. Li *et al.*

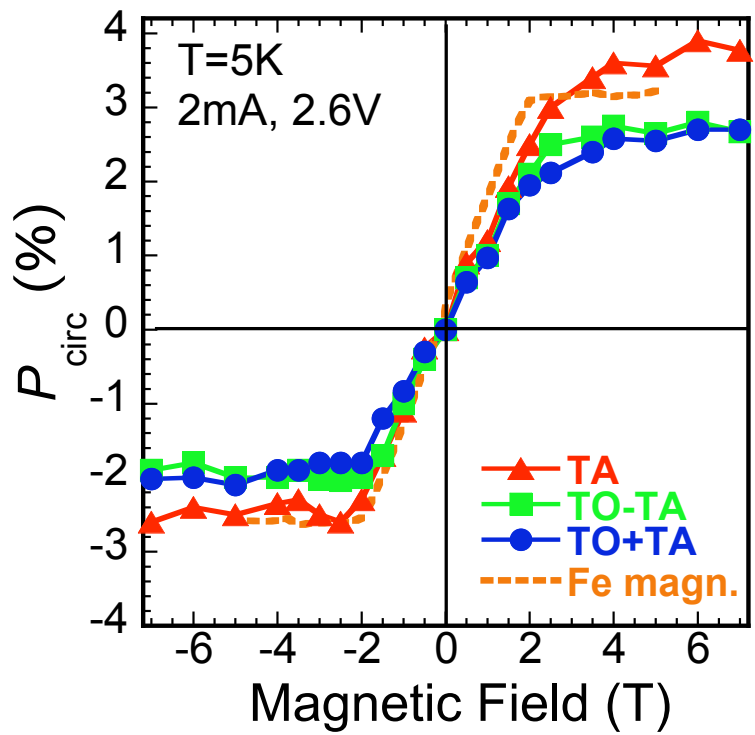

Figure 3. Li *et al.*

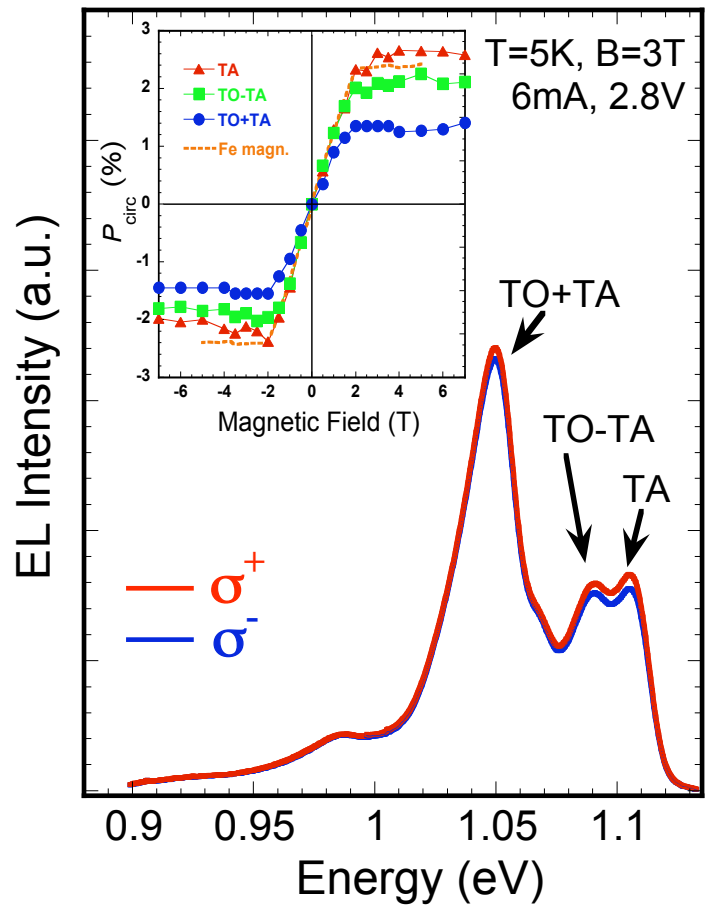

Figure 4. Li *et al.*